\documentclass[aps,prb,groupedaddress,showpacs,preprintnumbers,amsmath,amssymb,twocolumn,floats]{revtex4}

\usepackage{graphicx}
\usepackage[dvipdfm]{hyperref}
\usepackage{booktabs}
\usepackage{float}
\usepackage[]{sidecap}
\usepackage{CJK}
\usepackage{txfonts}

\begin{document}
\begin{CJK*}{UTF8}{gbsn}

\title{The design of a multi-channel spin polarimeter}

\author{ Tan Shi $^1$ }

\author{Fuhao Ji  $^1$}
\author{Mao Ye  $^2$ }
\author{Weishi Wan  $^3$ }
\author{Shan Qiao  $^2$ $^4$}\email{qiaoshan@mail.sim.ac.cn}

\affiliation{1. Physics Department, Laboratory of advanced Materials and Surface Physics Laboratory , Fudan University, Shanghai, 200438, China}
\affiliation{2.Statekey laboratory of functional materials for informatics, Shanghai Institute of Microsystem and Information Technology, Chinese Academy of Sciences, 865 Changning Road, Shanghai, 200050, China }
\affiliation{3.Advanced Light Source, Lawrence Berkeley National Laboratory, 1 Cyclotron Road,Berkeley, CA 94720, USA}
\affiliation{4.School of physical science and technology, ShanghaiTec University, 319 Yueyang Road,Shanghai 200031, China}

\date{\today}


\begin{abstract}

All commercial electron spin polarimeters work in single channel mode, which is the bottleneck of researches by spin-resolved photoelectron spectroscopy. By adopting the time inversion antisymmetry of magnetic field, we developed a multichannel spin polarimeter based on normal incident VLEED.  The key point to achieve the multi-channel measurements is the spatial resolution of the electron optics. The test of the electron optics shows that the designed spatial resolution can be achieved and an image type spin polarimeter with 100 times 100, totally ten thousand channels is possible to be realized.
\end{abstract}


\maketitle
\end{CJK*}

\section{Introduction}

Properties of materials are determined by electronic states. As the fast improvement of the power of computer, we can possibly predict the properties of a material from its electronic states. A typical example is the discovery of topological insulator whose properties are predicted by theory primarily. So the measurements of the electronic states of materials take the key role to understand the mechanism of their physical and chemical properties.

There are only three good quantum numbers, energy, momentum and spin to describe the electronic states in materials. The most direct method to study them is angle-resolved photoelectron spectroscopy (ARPES). In this method, from momentum and energy conservation laws, the electronic states inside materials can be determined by the measurements of energy and momentum of free photoelectrons. The classical ARPES performs this measurement by changing the voltages of the analyzer and the emission angle of photoelectrons by rotating the sample, so only the intensity of photoelectrons with a certain kinetic energy and a certain emission angle can be recorded at the same time and this measurement mode is called single channel one. The invention of image type multi-channel electron analyzer by K.Siegbahn, who won the Nobel Prize in 1981 because his contribution to high resolution photoelectron spectroscopy, results in the qualitative change of ARPES and makes the high-resolution measurements for both energy and momentum possible. The commercial electron analyzer can achieve the resolution of 0.6 meV and 0.1 degree for energy and momentum measurements due to the image type multi-channel mode.

Major discoveries of modern condensed matter physics, such as superconductivity, quantum Hall effect, Kondo effect, giant magnetic resistance, spin density wave, topological insulators et al., are all related with one or two of the strong correlation and spin-orbit interactions. Both interactions are closely related with the spin of electrons, so to study the mechanism of some novel physical properties, spin must be measured.

\begin{figure}
\includegraphics[width=8cm]{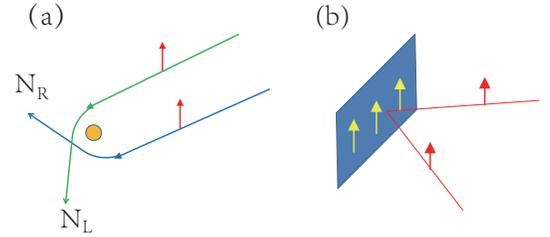}
\caption{(Color online) The mechanism of spin polarimeters based on spin-orbit(a) and strong correlation (b)interactions.
 }\label{fig1}
\end{figure}

\begin{figure*}
\includegraphics[width=18cm]{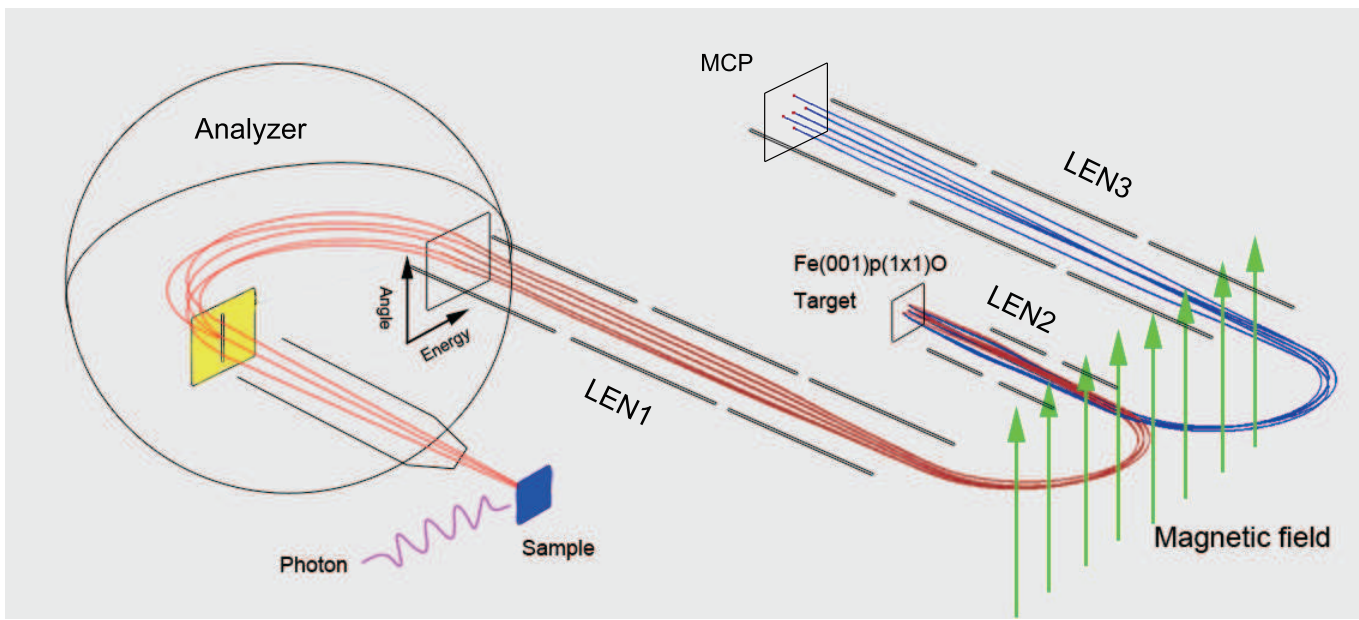}
\caption{(Color online) The sketch of the spin-resolved ARPES spectrometer based on multi-channel VLEED .
 }\label{fig3}
\end{figure*}
All the current available spin polarimeters also base on these two interactions mentioned above. Mott, spin low-energy-electron-diffraction (SLEED) and diffuse scattering polarimeters are based on spin-orbital interaction, which is a relativity effect that electric field in motion will induce magnetic field. A simplified explanation is shown in Fig.~\ref{fig1} (a). When polarized electron is scattered by a high Z atom, in the coordination system siting on electron, the charged nuclear of the atom will rotate around the electron and results in electric current and the current will generate magnetic field. The electrons scattered to left and right will see magnetic fields in up and down directions, respectively. The interactions between the magnetic fields and the spin of electrons have opposite sign and will result in different scattering intensities $N_L$ and $N_R$. The asymmetry A

\begin{eqnarray}
A &=& \frac{ N_{L}-N_{R}}{N_{L}+N_{R}}
\end{eqnarray}
is proportional to the spin polarization P of incident electrons,
\begin{eqnarray}
A &=& PS
\end{eqnarray}
where the ratio S is called Sherman function and is the scattering asymmetry of 100\% spin polarized electrons. The efficiency $\epsilon$ of a spin polarimeter is~\cite{lab1}
\begin{eqnarray}
\epsilon &=& \frac{I}{I_{0}}S^2
\end{eqnarray}
where I = $N_L$+$N_R$, is the total intensity of scattered electrons and $I_0$ is the total intensity of incident electrons. For all spin polarimeters based on spin-orbital interaction, the efficiency is about $1\times 10^{-4}$. Classical Mott spin polarimeter~\cite{lab2} works at 50$\sim$100 keV and the insulation between target and detectors is very difficult and the equipment is huge. Retarding type compact Mott polarimeter operating at several tens keV~\cite{lab3,lab4,lab5} is the most popular selection because its very stable performance. SLEED~\cite{lab6} and diffuse scattering~\cite{lab7} polarimeters work at low kinetic energy from tens to a hundred electronvolt and the performance is strongly dependent on the cleanness of target surfaces and periodic cleaning processes are necessary.  SLEED and diffuse scattering polarimeters use W(100) single crystal and Au film respectively as the scattering targets. Recently, very low energy electron diffraction (VLEED)~\cite{lab8} spin polarimeter has been invented which  utilizes strong correlation interaction. Its mechanism is shown in Fig.~\ref{fig1} (b). The ferromagnetic FeO target is used and the spin of electrons inside it is aligned after saturated magnetization. If the spin of incident electrons is parallel with that in target, from Pauli exclusion principle, the electrons cannot enter the target and a large scattering intensity will be observed compared with the antiparallel case. This scattering asymmetry is used to measure the spin polarization of incident electrons with an efficiency of about $10^{-2}$,  100 times higher than that based on spin-orbit interaction. Even for VLEED polarimeter, the efficiency is still poor, which makes spin-polarized ARPES  measurements a time consuming one.

Although the spin measurements are very important for modern material sciences, the current available commercial spin polarimeters all work in single channel mode. Many important fields remains inaccessible with current single channel spin polarimeter, because the low measuring efficiency limits the energy and angular resolution. To overcome this problem, many scientists try to develope multi- channel spin polarimeters. The key point is to design an electron optics with small aberrations to realize high spatial resolution. Until now, only one multi-channel spin polarimeter based on SLEED was invented by Kirschner group in Germany~\cite{lab9}. In their design, the incident electron reaches W(100) target with a $45^0$ incident angle and undergoes an aberration free specular reflection. A virtual image was created behind the W(100) target by  electron lens and the final real image is formed on the detector by  the specular reflection. A total of 1044 channels can be realized. However, the principle of this multi-channel spin polarimeter is based on spin-orbit interaction, whose efficiency is only one percent of VLEED's, and a new design based on VLEED is strongly required. Moreover, their polarimeter can only measure ferromagnetic samples that the spin polarization is measured by the asymmetry of scattering intensities of two success observations with the sample be magnetized in two antiparallel directions. The studies of novel phenomena in non-ferromagnetic materials related with spin-orbit interaction have become frontiers of modern condensed matter physics, for example, topological insulator, Rashiba effect et al., so the extension of the research object to non-ferromagnetic materials is an urgent task. Here, we report the design of a novel multi-channel spin polarimeter based on normal incident VLEED which can overcome the two shortcomings of SPLEED type mentioned above.

\begin{figure}
\includegraphics[width=8cm]{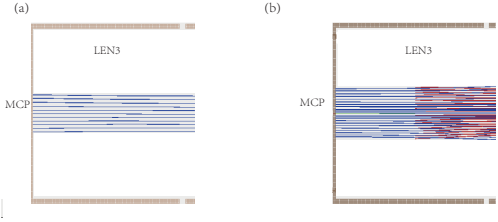}
\caption{(Color online)  Ray tracing results of the electron optics along energy (a) and momentum(b) directions.
 }\label{fig4}
\end{figure}

\section {Design of the multi-channel spin detector}

For VLEED polarimeter, the Sherman function strongly depends on incident angle and the maximum scattering asymmetry is obtained for normal incident~\cite{lab10}. In order to realize normal incident, the time reverse anti-symmetry of magnetic field is used. The structure of the normal incident multi-channel VLEED spin polarimeter is shown in part of Fig.~\ref{fig3}. The spin polarimeter consists of electron optics, target and detector. The detector is a microchannel plate (MCP) with fluorescence screen. The electron intensity image is recorded by a CCD camera behind the fluorescence screen. The function of electron optics is realized by three electron lenses and a dipole magnetic field. The electron beam from the same point on the incident plane of polarimeter becomes a parallel one after lens 1 (Len1), and turns 180 degree by the magnetic field, and then re-focus on the ferromagnetic FeO target by lens 2 (Len2). The normal incident geometry makes very small aberrations, which guarantees the good spatial resolution. After being scattered by the target, the electron beam becomes parallel one again after Len2 and re-enters the magnetic field. After turned 180 degree again, the beam finally focus on the entrance plane of MCP. Fig.~\ref{fig3} shows the whole diagram of our spin-resolved APERS spectrometer. A Scienta R3000 electron analyzer is adopted. The angle (momentum) and energy directions of electron analyzer are parallel with and perpendicular to the magnetic field, respectively. The length of the Len1, Len2 and Len3 are 359 mm, 133 mm and 393 mm, respectively. The radius of the electron orbit in the magnetic field is 60 mm. Our design makes the switch between the spin-resolved and the spin-integrated  modes very simple by only the change of the magnetic field. When we decrease the magnetic field to the half of the spin-resolved mode, the electron beam will turn a radius of 120 mm, go to Len3 directly and be focus on the MCP behind Len3. In this case, the spin-integrated, that is normal ARPES measurements can be performed.

\begin{figure}
\includegraphics[width=8cm]{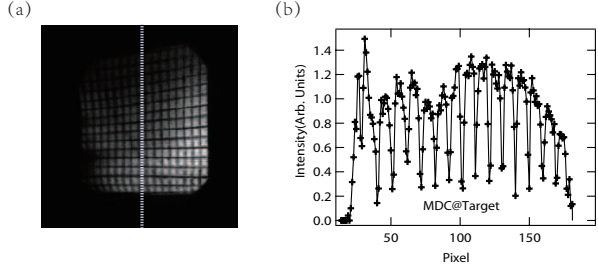}
\caption{(Color online)  (a)The electron intensity image on the target. (b)The intensity spectrum cut from (a) along momentum direction
 }\label{fig5}
\end{figure}

\begin{figure}
\includegraphics[width=8cm]{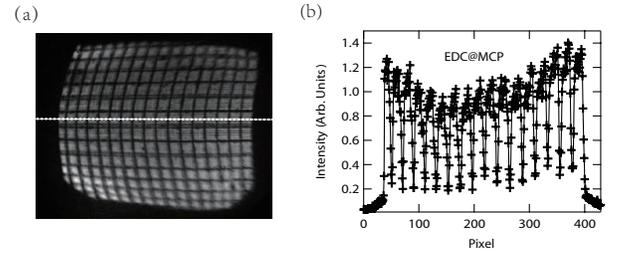}
\caption{(Color online)  (a)The electron intensity image on final MCP. (b)
The intensity spectrum cut from (a) along energy direction
 }\label{fig6}
\end{figure}

When an electron passes through magnetic field, the projection of electron spin along magnetic field direction is conserved, and the projection perpendicular to magnetic field undergoes a Larmor precession with a frequency
\begin{eqnarray}
w_{s} &=& \frac{eB}{m}.
\end{eqnarray}
On the other hand, the frequency $\omega_O$ of the electron orbital movement can be estimated to be the same as $\omega_s$ discussed below, so the projection of electron spin perpendicular to the magnetic field will turn just 180 degree and the Larmor procession does not destroy the function of the polarimeter. For electron move in magnetic field, the velocity v, orbital radius R and the intensity of magnetic field B have the relation
\begin{eqnarray}
evB &=& \frac{mv^2}{R}.
\end{eqnarray}
So,
\begin{eqnarray}
v &=& \frac{eBR}{m},
\end{eqnarray}
and the orbital frequency $w_0$
\begin{eqnarray}
w_{0}=\frac{v}{R}=\frac{eB}{m}=w_{s}.
\end{eqnarray}

The ray tracing results of the electron optics by SIMION program developed by Scientific Instrument Services, Inc.~\cite{lab11} are shown in Fig.~\ref{fig4} . The cross section of entrance plane of MCP is the leftmost vertical line. The beam sizes on the entrance plane of MCP in energy and momentum directions are smaller than 0.14 mm and 0.2 mm respectively. The effective areas of the polarimeter along these two directions are $\pm$ 8 mm and $\pm$ 10 mm respectively, so more than 100 channels can be achieved in both directions.

\section{Performance of the electron optics and VLEED target}

To test the performance of the electron optics, two meshes were set at the entrance plane of spin polarimeter which is the entrance plane of original MCP of electron analyzer that was removed when connected with the spin polarimeter. The diameters of metal wires of the meshes were 0.2 mm and 0.05 mm, respectively. An electron gun was set in the analyzer chamber and the elastically scattered electrons from sample were collected by the electron analyzer. A MCP was set just at the target position behind Len2 to test the performance of Len1, Len2 and magnetic field. The recorded electron intensity image is shown in Fig.~\ref{fig5} (a). The wires with 0.2 mm diameter can be clearly distinguished. We can see the spatial resolutions along momentum and energy directions are almost the same. To estimate the spatial resolution on target, the vertical slice is cut from Fig.~\ref{fig5} (a)  along the white broken line. After remove the smooth background, the intensity spectrum that is the momentum distribution curve(MDC) was obtained and shown in Fig.~\ref{fig5} (b).  The spatial resolution can be estimated from the broadening of peak edges. The spatial resolution should small than the number of pixel in which the edge height changes from 12\% to 88\%. From the MDC curve, the estimated average spatial resolution is smaller than 1.4 pixel. The total number of pixel is 163 times 163, so the spatial resolution of electron optics, which consists of Len1, Len2 and magnetic field, can guarantee at least 116 times 116 channels on the target.  Also we tested the performance of Len1, Len3 and magnetic field by the same method just mentioned. By decrease the magnetic field, the incident electrons directly hit the final MCP behind Len3 without passing through Len2. The electron intensity image on MCP is shown in Fig.~\ref{fig6} (a) and the energy distribution curve (EDC) after the remove of background from the horizontal slice is shown in Fig.~\ref{fig6} (b). Both the 0.2 mm and 0.05 mm wires can be clear seen. The estimated average spatial resolution is less than 2.42 pixel. The total number of pixel is 365, so at least 150 times 150 channels can be guaranteed.

\section{conclusion}

We have designed a normal incident image type multi-channel VLEED spin polarimeter. Both ray tracing calculation and performance test show the good performance of our electron optics and  100 times 100 channels can be achieved. The final commission of the whole system will be finished soon.

\end{document}